\documentclass[preprint,12pt,authoryear]{elsarticle}
\usepackage[T1]{fontenc}
\usepackage[utf8]{inputenc}
\usepackage[english]{babel}
\usepackage{amsmath,booktabs,graphicx,float}
\usepackage[hidelinks]{hyperref}
\biboptions{round}
\journal{Oxford Bulletin of Economics and Statistics}
\makeatletter
\def\ps@pprintTitle{%
  \let\@oddhead\@empty\let\@evenhead\@empty
  \def\@oddfoot{\hfil\thepage\hfil}\let\@evenfoot\@oddfoot}
\let\PixDefaultCaption\@makecaption
\def\PixTableType{table}
\long\def\@makecaption#1#2{%
  \ifx\@captype\PixTableType
    \vskip\abovecaptionskip
    {\centering\MakeUppercase{#1}\par\textit{#2}\par}%
    \vskip\belowcaptionskip
  \else
    \PixDefaultCaption{#1}{#2}%
  \fi}
\makeatother
\abstracttitle{Abstract}
\keywordtitle{Keywords}
\newcommand{\ResultRows}{
\multicolumn{6}{@{}l}{\textit{A. Transaction count}} \\
3 & S1 & 608.60 & 652.38 & $-6.71$ & $[-15.29;\,0.54]$ \\
3 & S2 & 608.60 & 627.81 & $-3.06$ & $[-11.09;\,4.60]$ \\
3 & S3 & 608.60 & 623.39 & $-2.37$ & $[-9.87;\,5.48]$ \\
37 & S1 & 8639.48 & 9429.06 & $-8.37$ & $[-13.96;\,-4.53]$ \\
37 & S2 & 8639.48 & 9010.72 & $-4.12$ & $[-8.62;\,0.04]$ \\
37 & S3 & 8639.48 & 8976.32 & $-3.75$ & $[-8.49;\,0.76]$ \\
\addlinespace
\multicolumn{6}{@{}l}{\textit{B. Settlement value}} \\
3 & S1 & 183.72 & 216.60 & $-15.18$ & $[-32.17;\,2.98]$ \\
3 & S2 & 183.72 & 203.15 & $-9.56$ & $[-28.47;\,7.37]$ \\
3 & S3 & 183.72 & 206.17 & $-10.89$ & $[-29.69;\,6.57]$ \\
37 & S1 & 3787.32 & 4247.34 & $-10.83$ & $[-18.28;\,-3.83]$ \\
37 & S2 & 3787.32 & 3931.81 & $-3.68$ & $[-11.38;\,3.25]$ \\
37 & S3 & 3787.32 & 3932.12 & $-3.68$ & $[-11.68;\,3.18]$ \\
\addlinespace
}

\newcommand{\DiagnosticRows}{
\multicolumn{4}{@{}l}{\textit{A. Transaction count}} \\
Three-day validation windows & 29 & 29 & 29 \\
Three-day MAPE (\%) & 2.81 & 2.64 & 2.63 \\
Three-day bias (\%) & 2.26 & -1.07 & -1.04 \\
37-day validation windows & 24 & 24 & 24 \\
37-day MAPE (\%) & 4.49 & 0.94 & 1.05 \\
37-day bias (\%) & 4.49 & -0.24 & -0.36 \\
BIC & $-2863.27$ & $-2898.82$ & $-2906.26$ \\
Ljung--Box $p$, lag 7 & $<0.001$ & $<0.001$ & $0.229$ \\
Ljung--Box $p$, lag 14 & $<0.001$ & $<0.001$ & $0.204$ \\
Ljung--Box $p$, lag 28 & $<0.001$ & $<0.001$ & $0.580$ \\
Ljung--Box $p$, lag 35 & $<0.001$ & $<0.001$ & $<0.001$ \\
\addlinespace
\multicolumn{4}{@{}l}{\textit{B. Settlement value}} \\
Three-day validation windows & 29 & 29 & 29 \\
Three-day MAPE (\%) & 6.79 & 5.39 & 5.38 \\
Three-day bias (\%) & 5.17 & 0.63 & 0.56 \\
37-day validation windows & 24 & 24 & 24 \\
37-day MAPE (\%) & 4.64 & 1.18 & 1.20 \\
37-day bias (\%) & 4.60 & -0.63 & -0.64 \\
BIC & $-830.12$ & $-850.02$ & $-823.60$ \\
Ljung--Box $p$, lag 7 & $0.165$ & $0.549$ & $0.173$ \\
Ljung--Box $p$, lag 14 & $0.002$ & $0.122$ & $0.022$ \\
Ljung--Box $p$, lag 28 & $0.004$ & $0.252$ & $0.042$ \\
Ljung--Box $p$, lag 35 & $0.003$ & $0.109$ & $0.012$ \\
\addlinespace
}

\begin{document}
\renewcommand{\figurename}{Figure}
\renewcommand{\tablename}{Table}
\begin{frontmatter}
\title{Instant Payments After a Betting Ban: Evidence from Brazil}

\author[santanna]{Rafael Voigtel\corref{cor1}}
\ead{r.voigtelcesar@studenti.unipi.it}
\author[columbia]{Guilherme Vianna}
\ead{gd2801@columbia.edu}
\address[santanna]{Sant'Anna School of Advanced Studies, Piazza Martiri della Libertà 33, 56127 Pisa, Italy}
\address[columbia]{Columbia University, Department of Economics, 1022 International Affairs Building, Mail Code 3308, 420 West 118th Street, New York, NY 10027, USA}
\cortext[cor1]{Corresponding author.}

\begin{abstract}
Brazil has been facing major problems with delinquency, household debt, high interest rates, consumption, and access to credit, while online sports betting has emerged in public debate as a possible contributing factor, culminating with the federal government banned betting nationwide On September 25, 2026. We compare the number and total value of instant payments (\textit{Pix}) over the following 37 days with forecasts based on pre-ban data, finding that both outcomes were below forecast in all three models. Under the two quadratic-trend models, the number of transactions was 3.75--4.12\% below forecast and total transaction value was about 3.68\% below forecast, while the linear-trend model gave larger gaps of 8.37\% and 10.83\%, respectively. These gaps offer indirect evidence on the scale of Pix flows potentially linked to betting.
\end{abstract}
\end{frontmatter}

\section{Introduction}

Brazilian households continue to face financial difficulties despite recent gains in employment and labour income. The Central Bank's September 2026 report described weaker household consumption and expensive credit, even as real labour income remained higher than a year earlier. Monetary policy remains restrictive, with the Selic rate at 13.75\% per year \citep{rpm}. In August, household loan delinquency, measured as arrears of more than 90 days, reached 6.04\%. The seasonally adjusted debt-service ratio was 28.99\%, and revolving credit-card interest averaged 15.18\% per month \citep{sgs}. Even when incomes rise, debt payments can leave households with little room to spend. The government made debt renegotiation a priority through \textit{Desenrola Brasil}, launched in 2023, and \textit{Novo Desenrola Brasil}\footnote{Both are federal government programs that renegotiate the debts of households and businesses.}, launched in 2026 \citep{desenrola2023,desenrola2026}.

Betting became part of the debate over household debt and social harm as Brazil expanded its legal betting market. Law No.\ 13,756 authorized fixed-odds betting in December 2018, and Law No.\ 14,790 extended the framework to online games in December 2023. The federally authorized market began operating on January 1, 2025 \citep{history}. Following Supreme Court orders, the government published rules in October 2025 restricting participation by beneficiaries of \textit{Bolsa Família} and \textit{Benefício de Prestação Continuada} (BPC) \citep{beneficiaries}. On September 25, 2026, Provisional Measure No.\ 1,394 banned fixed-odds betting on sporting events and online games nationwide. It preserved other legally authorized lottery categories and provided for closures and refunds during a transition. The explanatory memorandum cited debt, health risks and bettor vulnerability as reasons for the ban \citep{ban}.

Previous studies have examined how gambling affects spending, savings and other outcomes. \citet{kearney} links the introduction of state lotteries to lower spending on goods and services other than gambling, while \citet{guryan} find that increases in lottery spending persist, providing evidence of addiction. \citet{cotti} find that after casinos open, fatal alcohol-related traffic accidents increase in less populous counties and decrease in more populous ones. \citet{baker} find that savings fell after US states legalized online sports betting, with the decline concentrated among frequent bettors and households with little savings. Here we compare the number and total value of Pix transactions after the ban with forecasts based on data from before the ban. The daily data let us measure the size of the gap and trace how it evolved after publication.

\section{Data and methodology}

We study the daily number and total value of Pix transactions settled through the Instant Payments System (SPI). The data contain 2,190 daily observations from November 3, 2020 to November 1, 2026 \citep{spi}. We estimate the models using all 998 days from January 1, 2024 to September 24, 2026, including weekends and holidays. We convert total transaction value from thousands of reais to nominal reais, take natural logarithms of both outcomes and fit separate models for each.

We use the Causal-ARIMA approach of \citet{menchetti} to forecast what the number and total value of transactions would have been without the ban. We fit a regression with ARIMA errors to data from before the ban, allowing regression errors on different dates to be correlated.
\begin{equation}
\begin{aligned}
\log Y_t={}&\alpha+\beta_1\tau_t+\beta_2\tau_t^2
 +\sum_{d=2}^{7}\gamma_d D_{dt}+\sum_{m=2}^{31}\delta_m M_{mt}\\
&+\mathbf H_t'\boldsymbol\eta
 +\sum_{k=1}^{2}\left[a_k\sin(2\pi k\tau_t)+b_k\cos(2\pi k\tau_t)\right]+u_t
\end{aligned}
\end{equation}
Here, \(Y_t\) is the number of transactions or total transaction value, and \(\tau_t\) is elapsed time in years of 365.25 days. The indicators \(D_{dt}\) and \(M_{mt}\) account for the day of the week and the day of the month, with Monday and the first day of the month as the reference categories. The eight controls in \(\mathbf H_t\) cover holidays, the days before and after them, Christmas, New Year's Day, their eves and Ash Wednesday. Two pairs of sine and cosine terms account for annual seasonality, and we estimate the time trend jointly with the 48 calendar controls.

S1 sets \(\beta_2=0\), giving a linear trend in the log outcome, and uses weekly seasonal ARIMA\((1,0,0)(1,0,1)_7\) errors. To allow the growth rate to rise or fall over time, S2 adds a quadratic term and uses the same weekly seasonal ARIMA errors. For both models, \((1-\phi L)(1-\Phi L^7)u_t=(1+\Theta L^7)\varepsilon_t\), where \(L\) is the lag operator. S3 keeps the quadratic trend and uses \((1-\sum_{r=1}^{6}\phi_rL^r)u_t=\varepsilon_t\), an ARIMA\((6,0,0)\) process. The autocorrelation plots showed dependence especially at lags 3 and 6, motivating separate coefficients for the first six lags. S1 and S2 differ in their time trends, whereas S2 and S3 differ in how they model serial correlation in the regression errors.

We chose the ARIMA orders using model diagnostics during development. At each weekly cutoff, we kept the orders fixed and re-estimated the coefficients using only data available on that date. The comparison is retrospective because we examined historical forecast performance when developing the models.

We estimate the regression and ARIMA parameters jointly by Gaussian maximum likelihood. Forecasts on the original scale are \(\widehat Y_t^0=\exp(\widehat\mu_t+\widehat v_t/2)\), where the mean and predictive variance are on the log scale. For both outcomes, we compare totals over September 26--28 and September 26--November 1. Forecasts begin on September 25, but we exclude that day from the totals because it includes transactions made both before and after the measure was published. The longer period covers closures and refunds.

We calculate the percentage gap between the observed and forecast totals over period \(H\) as
\begin{equation}
\widehat\Delta_H=100\left(\frac{\sum_{t\in H}Y_t}{\sum_{t\in H}\widehat Y_t^0}-1\right)
\end{equation}
A causal interpretation of these gaps assumes a reliable no-ban forecast, no anticipation and no concurrent events that explain the gaps.

We use the \textit{stationary bootstrap} to construct approximate prediction intervals \citep{politis}.\footnote{Uncertainty in the three-day and 37-day totals depends partly on how forecast errors move together across days. Resampling blocks of consecutive days retains some of the dependence that may remain in the estimated one-step-ahead forecast errors. Re-estimating the model on each simulated pre-ban sample lets the intervals reflect uncertainty in the estimated coefficients.} We subtract the mean from the estimated one-step-ahead forecast errors (the innovations) and resample them in blocks of consecutive days. The block lengths follow a geometric distribution with a mean of 14 days and wrap back to the start if they reach the end of the series. In each replication, we simulate a pre-ban sample and re-estimate the model. We then use the observed data through September 24 and a separate draw of blocks to simulate the forecasts. We convert each simulated path to the original scale before summing its daily values and calculate the percentage gap using the simulated total in the denominator. The interval endpoints are the 2.5th and 97.5th percentiles of these gaps across 5,000 replications for each model and outcome.

\section{Results}

The number of transactions was 2.37\% to 6.71\% below forecast over three days and 3.75\% to 8.37\% below forecast over 37 days across the three models (Table~\ref{tab:results}). Total transaction value was 9.56\% to 15.18\% below forecast over three days; over 37 days, the percentage gap was \( -10.83\% \) under the linear-trend model (S1) and approximately \( -3.68\% \) under each quadratic-trend model (S2 and S3).

\begin{table}[H]
\centering\small\linespread{1}\selectfont
\caption{Cumulative deviations in transaction counts and settlement values}
\label{tab:results}
\setlength{\tabcolsep}{4pt}
\begin{tabular}{@{}llrrrr@{}}
\toprule
\textit{Days} & \textit{Model} & \textit{Observed} & \textit{Counterfactual} & \textit{Gap (\%)} & \textit{95\% interval} \\
\midrule
\ResultRows
\bottomrule
\end{tabular}
\par\smallskip
\begin{minipage}{0.98\linewidth}\footnotesize
Panel A is in millions of transactions. Panel B is in billions of nominal BRL.
Both windows start on September 26, 2026. Intervals use 5,000 valid stationary
bootstrap replicates per model and outcome. Data source: Central Bank of Brazil.
\end{minipage}
\end{table}

\begin{figure}[H]
\centering
\includegraphics[width=\linewidth]{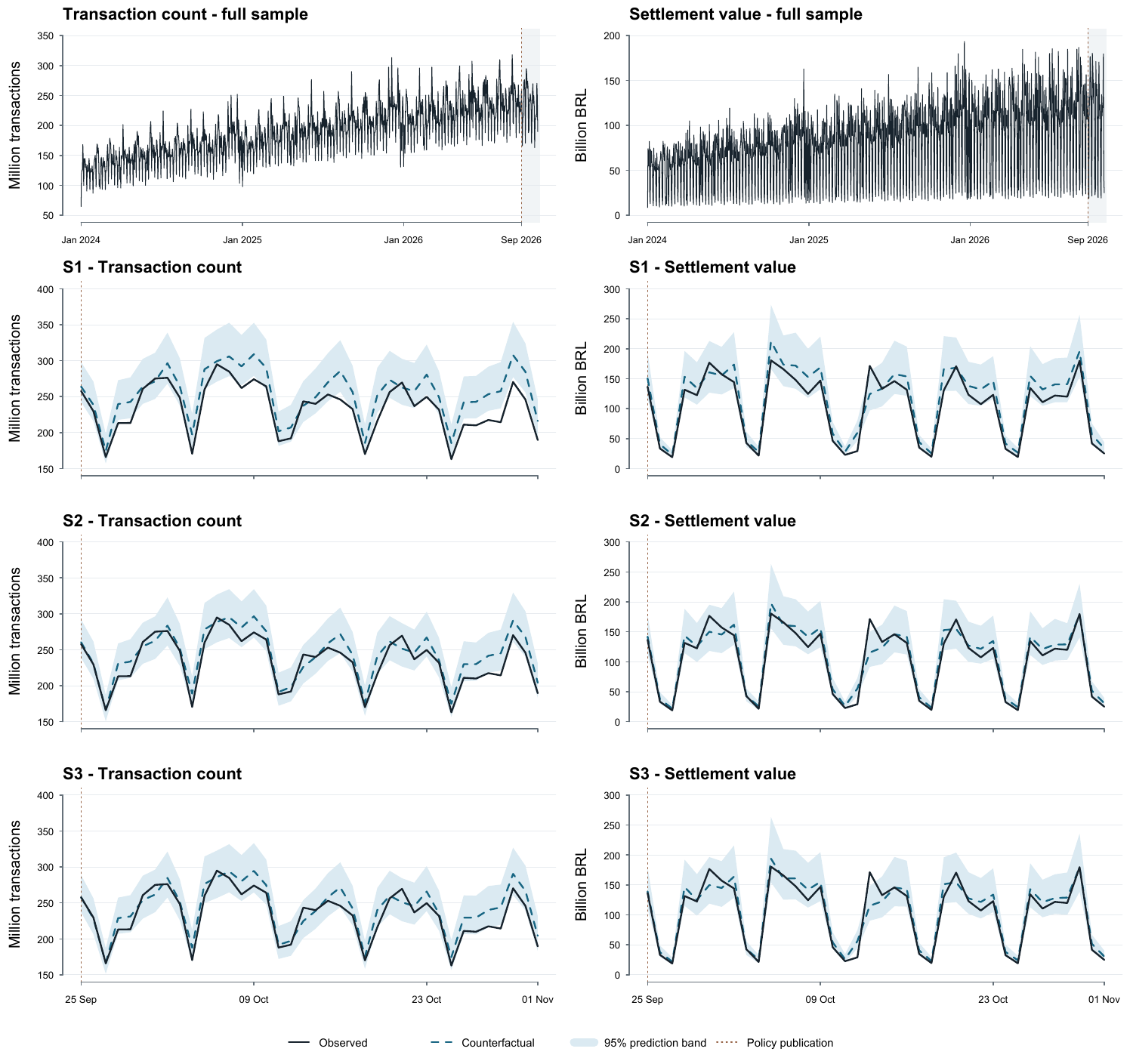}
\caption{Observed and counterfactual Pix activity. Columns show transaction counts and settlement values. The top row shows January 1, 2024--November 1, 2026, including all 998 estimation days through September 24, 2026. Lower rows detail forecasts for September 25--November 1, 2026. Shaded bands are pointwise 95\% prediction intervals. Vertical dotted lines mark policy publication on September 25. Data source: Central Bank of Brazil.}
\label{fig:both}
\end{figure}

Over 37 days, total transaction value was BRL~460 billion below the S1 forecast and approximately BRL~145 billion below each of the S2 and S3 forecasts. These gaps cover all Pix transactions settled through SPI, with all three models putting both outcomes below forecast. The 37-day intervals from the linear-trend model exclude zero, whereas every three-day interval and both quadratic-trend models' 37-day intervals include zero. Figure~\ref{fig:both} shows the full estimation sample and the forecasts for the period after the ban.

\section{Robustness and diagnostics}

We assess how well the models forecast before the ban by fitting each model once a week using the data available at each cutoff \citep{tashman}. For three-day forecasts, we compare forecast and observed totals for days 2--4 after the cutoff. For 37-day forecasts, we use days 2--38. Both exercises skip the first day after the cutoff, just as the main analysis skips the publication day. For the 37-day exercise, we use 84 weekly cutoffs before successive Fridays from January 3, 2025 to August 7, 2026. All 37-day windows end by September 13, 2026. We distinguish the first 60 cutoffs from the 24 most recent. Table~\ref{tab:diagnostics} reports MAPE and bias for the 24 recent 37-day windows and the 29 recent three-day windows for each outcome.

For each historical window, we calculate the percentage forecast error as \(100\left(\frac{\text{forecast total}}{\text{observed total}}-1\right)\). The mean error is labelled bias in Table~\ref{tab:diagnostics}. A positive mean indicates overprediction. The mean absolute percentage error (MAPE) averages the absolute errors. Over 37 days, the recent MAPE for transaction counts was 4.49\% under S1 and 0.94--1.05\% under S2 and S3. For transaction value, it was 4.64\% under S1 and 1.18--1.20\% under S2 and S3. S1 tended to overpredict both totals by about 4.5\%, whereas the absolute mean errors of S2 and S3 were below 0.65\%. In the earlier 60 windows, S1 had the lowest MAPE for both outcomes. The advantage of the quadratic trends was therefore specific to the more recent period.

\begin{table}[H]
\centering\small\linespread{1}\selectfont
\caption{Recent historical forecast accuracy and final-sample diagnostics}
\label{tab:diagnostics}
\begin{tabular}{@{}lrrr@{}}
\toprule
\textit{Measure} & \textit{S1} & \textit{S2} & \textit{S3} \\
\midrule
\DiagnosticRows
\bottomrule
\end{tabular}
\par\smallskip
\begin{minipage}{0.98\linewidth}\footnotesize
Positive bias indicates overprediction. The three-day windows end by September 17, 2026. The 37-day windows end by September 13, 2026.
Weekly 37-day windows overlap. Ljung--Box entries are conventional \(p\)-values.
BIC comparisons are made within each outcome.
\end{minipage}
\end{table}

The diagnostics use models estimated on the full pre-ban sample. The Bayesian information criterion (BIC), which weighs model fit against complexity, favours S3 for the number of transactions and S2 for total transaction value. Each comparison is made within an outcome. The Ljung--Box tests jointly assess residual autocorrelation through maximum lags of 7, 14, 28 and 35 days, corresponding to one, two, four and five weeks. Because the tests use residuals from before the ban, the same results apply to both post-ban periods. For total transaction value, S2 is the only model for which all four tests fail to reject the null of no autocorrelation at the 5\% level. For the number of transactions, the tests for S3 reject the null only when they include lags through day 35.

We also use these forecasts to compare the observed post-ban gaps with the gaps on the 84 earlier dates. As in the main analysis, each percentage gap uses the forecast total as its denominator. None of the 24 recent windows had a gap as negative as the post-ban gap for any model or outcome. Across all 84 windows, the counts were 2, 3 and 4 for transaction counts and 3, 7 and 8 for transaction value (S1 to S3). The comparison describes how often gaps of this size arose before the ban. Because the windows overlap, these frequencies should be read as descriptive comparisons.\footnote{Online Table S1 and Figure S1 report the full placebo comparison. Online Table S2 reports forecast accuracy for the two historical periods and for all 84 windows.}

The endpoints of the interval for the 37-day percentage gap in total transaction value changed little between 1,000 and 5,000 replications.

\section{Conclusion}

The number and total value of Pix transactions were below forecast after the betting ban in all specifications considered, with statistical significance varying between chosen models. Over 37 days, the linear-trend model (S1) forecast 789.6 million more transactions and BRL~460.0 billion more in total transaction value than observed, with observed totals 8.37\% and 10.83\% below those forecasts, respectively. The two quadratic-trend models gave smaller gaps of 336.8--371.2 million transactions and BRL~144.5--144.8 billion in total transaction value. Using daily payment data, the analysis quantifies the size and evolution of these gaps following the nationwide ban and provides indirect evidence on the scale of Pix flows potentially linked to betting.

\section*{Funding}
This study received no funding.

\section*{Data availability}
The SPI series and SGS indicators are publicly available from the Central Bank of Brazil. The archived input workbook and the R code will be available.

\section*{Declaration of generative AI use}

During the preparation of this manuscript, the authors used ChatGPT (OpenAI) to assist with developing and reviewing the R code and translating and editing the manuscript in English. The authors reviewed and edited the generated material and take full responsibility for the code, analysis and final content.

\begingroup
\raggedright

\endgroup
\end{document}


\begin{center}
{\Large\bfseries Online supplement\par}
\medskip
{\large Instant Payments After a Betting Ban: Evidence from Brazil\par}
\end{center}

For each of the 84 Fridays from January 3, 2025 to August 7, 2026, we estimate each model with data available through the previous day. We compare forecast and observed totals from the Saturday after each Friday through day 38 after the estimation cutoff. This gives 37 days in each comparison, as for the September 25, 2026 policy date. The earlier period comprises the first 60 windows and the recent period the last 24.

\begin{table}[H]
\centering\footnotesize
\caption{Post-ban gaps and the distribution of earlier placebo gaps}
\label{tab:placebo}
\setlength{\tabcolsep}{3pt}
\resizebox{\linewidth}{!}{%
\begin{tabular}{@{}lllrrrrrrr@{}}
\toprule
 & & & & & \multicolumn{2}{c}{As low as actual} & \multicolumn{3}{c}{Historical quantiles (\%)}\\
\cmidrule(lr){6-7}\cmidrule(l){8-10}
Outcome & Model & Period & Actual (\%) & $N$ & Count & Share (\%) & 2.5th & 50th & 97.5th\\
\midrule
\PlaceboRows
\bottomrule
\end{tabular}}
\par\smallskip
\begin{minipage}{\linewidth}\footnotesize
\textit{Notes.} A gap is the observed total divided by the forecast total, minus one, expressed as a percentage. A placebo counts as low if its gap is at or below the post-ban gap for the same model and outcome. The historical windows overlap. The shares describe the sample, and the historical quantiles describe its spread.
\end{minipage}
\end{table}

\begin{figure}[p]
\centering
\includegraphics[width=\linewidth]{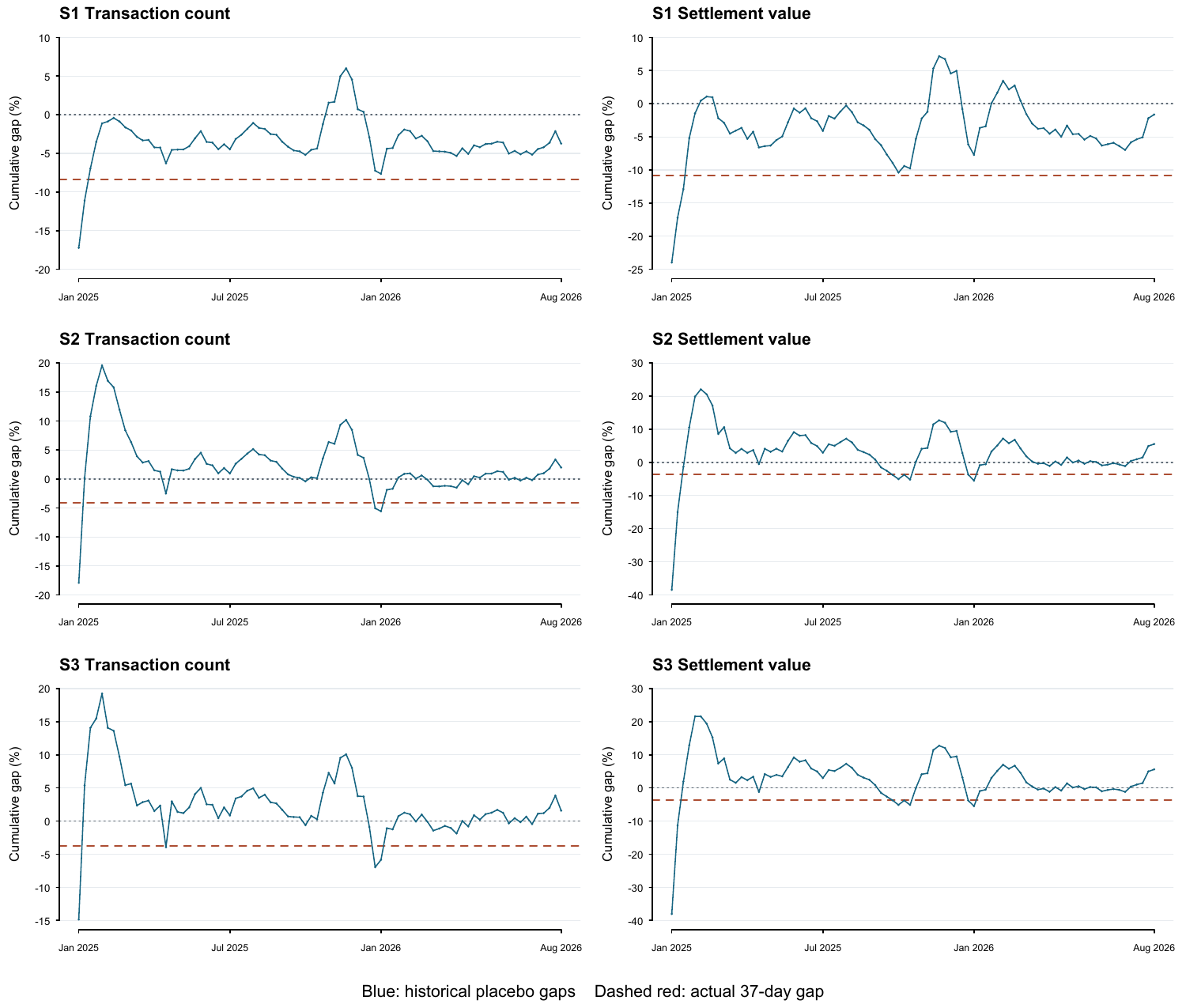}
\caption{Cumulative gaps for the 84 placebo dates. The dashed line in each panel is the actual post-ban gap for that outcome and model. The vertical scale varies across panels}
\label{fig:placebo}
\end{figure}
\clearpage

\begin{table}[H]
\centering\small
\caption{Forecast accuracy over 37 days by historical period}
\label{tab:validation}
\begin{tabular}{@{}lllrrr@{}}
\toprule
Outcome & Model & Period & Windows & MAPE (\%) & Bias (\%)\\
\midrule
\ValidationRows
\bottomrule
\end{tabular}
\par\smallskip
\begin{minipage}{\linewidth}\footnotesize
\textit{Notes.} The forecast error is the forecast total divided by the observed total, minus one, expressed as a percentage. MAPE is the mean absolute error in those percentage terms. Bias is the mean signed error, with positive values indicating overprediction. All results use Central Bank of Brazil data.
\end{minipage}
\end{table}